\documentclass[twocolumn,showpacs,amsmath,amssymb,superscriptaddress,
 aps,
 prb,
 floatfix,
 lengthcheck,
]{revtex4-1}

\usepackage{graphicx}
\usepackage[english]{babel} 
\usepackage[utf8]{inputenc}
\usepackage{hyperref}
\usepackage{color}
\usepackage[normalem]{ulem}
\usepackage{xcolor}

\newcommand{\cF}{\ensuremath{\mathcal F}}
 
\begin{document}

\title{
	Current fluctuations of diffusive systems with a battery
}

\author{Thibaut Jonckheere}
\affiliation{Aix Marseille Univ, Universit\'e de Toulon, CNRS, CPT, Marseille, France}
\email{thibaut.jonckheere@cpt.univ-mrs.fr }
\author{Bernard Derrida}
\affiliation{Coll\`ege de France, 11 place Marcelin Berthelot, 75005 Paris, France}
\affiliation{Laboratoire de Physique de l'Ecole Normale Sup\'erieure, ENS, Universit\'e PSL,  CNRS, Sorbonne Universit\' e, Universit\' e de Paris, F-75005 Paris, France}
\email{derrida@phys.ens.psl.eu}

\date{\today}

\begin{abstract}
For a diffusive system  on a ring, with a battery at the origin, we use the macroscopic fluctuation theory (MFT) to obtain the large deviation function of the current. In the case of non-interacting particles, the predictions of the MFT agree with the result of a direct exact microscopic calculation. To our surprise,  the large deviation function of non-interacting particles with a battery has the same form as for the Symmetric Simple Exclusion Process (SSEP) with open boundary conditions.
For the SSEP on a ring with a battery, the predictions of the MFT seem to agree with the results of long Monte Carlo simulations,  although much longer simulations would be needed to better test the validity of these predictions.
\end{abstract}


	\maketitle

		\section{Introduction}
		Understanding the properties of  steady states is one of the central questions in the study of  non-equilibrium statistical physics.
		One way of maintaining  a system of particles  in a non-equilibrium steady state is to connect it to two or more  reservoirs at unequal densities (or equivalently at unequal chemical potentials).\cite{BDGJL2,BDGJL6,Bodineau1,Derrida 2007}

Alternatively one can impose a non-conservative force, 
		for example by a battery which fixes a jump of the chemical potential at a given position (as in Ref.~\onlinecite{battery} where
		the two-point correlations where computed in the steady-state).   
In this case, 
if there is no exchange of particles with the outside world,
 the number of particles in the system is  fixed. 
		In the present work we try to develop a theory to characterize the fluctuations and the large deviations of the current 
of diffusive systems
in presence of a battery, which can be thought as a very localized non-conservative force \cite{battery,Sadhu}.

        The macroscopic fluctuation theory (MFT) \cite{BDGJL1,BDGJL2,BDGJL6,Derrida 2007,Les Houches,Espigares_Hurtado-Garrido-PRL-(2013), Hurtado-(2025)} has been widely used to compute the fluctuations and the large deviations  of the current of one dimensional diffusive systems in contact with two reservoirs \cite{Bodineau1,BDGJL-2005-PRL, BDGJL-2006-JSP},
  on a ring \cite{Appert,Bodineau-Derrida-2005,Zarfaty},  on the infinite line  \cite{Antoine,Berlioz,Bet1,Krapivsky,Mallick,Gra1} or in higher dimensions \cite{Akkermans-Bodineau-Derrida-Shpielberg,Bodineau-2026}

Here under some assumptions  discussed in Appendix \ref{MFT} (Eq.~(\ref{ZZ5}) that the optimal profile for a large deviation of the current is time-independent, and Eq.~(\ref{ZZ6}) that the battery imposes a fixed relation between the densities at its boundaries)  we obtain, using the MFT,  a parametric expression of the large deviation of the current Eqs.~(\ref{BB6})-(\ref{BB8}) allowing to
compute the cumulants of the current on a ring with a localized battery.

We compare  these predictions  with a direct exact calculation in the case of non interacting particles
and with numerical Monte Carlo simulations in the case of the symmetric simple exclusion process SSEP.

 The paper is organized as follows. In section~\ref{sec:noninteracting}, we consider the case of non-interacting particles on a ring with a battery,
	where we derive the steady-state profile, and the current distribution. We observe
 in particular the surprising fact that this current distribution is similar to the one
	of  the symmetric simple exclusion process (SSEP)  for open-boundary conditions.
	 In section~\ref{sec:MFT1}, we apply the macroscopic fluctuation theory (MFT) to the case of a  general diffusive system on a ring  with a battery, 
	 which allows us to obtain analytic predictions for the  current along the ring and its 
	fluctuations. 
	 We then consider in section~\ref{sec:SSEP}  the symmetric simple exclusion process (SSEP) with a battery, and we compare numerical results obtained 
	using Monte-Carlo simulations with predictions of the MFT.  Finally, section~\ref{sec:conclusion} is devoted to concluding remarks and discussions. Details of the formalism and of the calculations are given in appendices.

{\section{Non interacting particles on a ring with a battery}
\label{sec:noninteracting}
	In this section, we solve explicitly the case of $N$ {\it non-interacting} particles on a ring of $L$ sites with a battery between site $L$ and site $1$
	(see Fig.\ref{fig:anneau} for a schematic view of the system). 
  Each particle on site $1\le i  \le L-1$ jumps to site $i+1$ at rate $1$ and similarly each particle  on site $2 \le i \le L$ jumps to site $i-1$ at rate $1$. The effect of the battery  localized between site $L$ and site $1$ is that a particle on site $L$  jumps at rate $G$ to site $1$ and a particle on  site $1$ jumps to site $L$ at rate $H$.
To our surprise   we will see that  the large deviation function of the current for non-interacting particles turns out to have a similar form  as in the case of  the SSEP on an open interval. We will also see that
		non-interacting particles on a ring with a battery exhibit  long-range correlations 
 in contrast to what they do for open boundary conditions.
 \begin{figure}
\centerline{\includegraphics[width=8.cm]{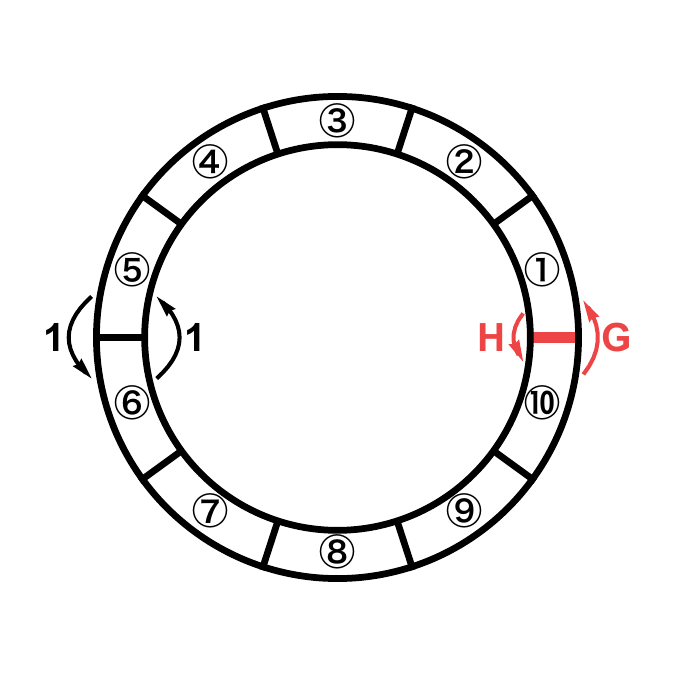} }
\caption{ Schematics of a ring with $L=10$ sites. The transfer rate through standard links (black) is 1 for both transfer directions. There is one special link (red, thick) between sites $L$ and $1$, with transfer rate $G$ from site $L$ to site $1$, and  transfer rate $H \leq G$ from site $1$ to site $L$. This special link plays the role of a battery.}
\label{fig:anneau}
\end{figure}
\subsection{The steady state profile and the long-range correlations}

			   Choosing $G \neq H$ induces a current and starting from any initial configuration, the system reaches, in the long time limit,  a steady state.

			   If $n_i(t)$ is the number of particles  on site site $i$, 
		   it is easy to see that  the average occupations $\langle n_i \rangle$ evolve according to
			   \begin{align*}
	{d \langle n_1 \rangle \over dt} &=  G \langle n_{L} \rangle + \langle n_2 \rangle -(1+H) \langle n_1 \rangle   &  \\
	{d \langle n_i \rangle \over dt} & =   \langle n_{i-1} \rangle + \langle n_{i+1} \rangle -2 \langle n_{i} \rangle  \ \ \  \ \ \text{for} \ \ \ 2 \le i \le L-1  \\
	{d \langle n_L \rangle \over dt} &=   \langle n_{L-1} \rangle + H \langle n_{1} \rangle -(1+G) \langle n_{L} \rangle    \ . 
	\end{align*} 
	This leads to the following steady state values
		\begin{equation}
	\langle n_i \rangle = 
	{ 2 \,N \, \big(1+ G  L  - H -i (G-H) \big) \over  L \, \big(2 +   (G+H) (L-1))\big) }
	\label{AA1}
	\end{equation}
	{ The calculation can be extended to the two-point correlations.
Let $y_k^{(i)}$ be equal to one if particle $k$ occupies site $i$,
and zero otherwise,
so that $n_i=\sum_{k=1}^N y_k^{(i)}$, and
$\langle y_k^{(i)}\rangle=p_i$ (where $p_i$ is the probability of finding a given particle on site $i$). Since different particles are independent,
whereas a given particle can occupy only one site, one has
\begin{equation}
\left\langle y_k^{(i)} y_{k'}^{(j)}\right\rangle
=
\begin{cases}
p_i p_j, & k\neq k',\\
p_i\delta_{ij}, & k=k'.
\end{cases}
\end{equation}
It follows that
\begin{equation}
\langle n_i n_j\rangle=N(N-1)p_i p_j+N p_i\delta_{i,j}.
\end{equation}
Using $\langle n_i\rangle=Np_i$, the connected correlation is therefore
		\begin{equation}
	\langle n_i n_j \rangle_c = 
		\langle n_i n_j \rangle  -
		\langle n_i\rangle \langle n_j\rangle  =  \langle n_i \rangle \, \delta_{i,j}  -
		{\langle n_i\rangle \langle n_j\rangle   \over N}
	\label{AA2}
	\end{equation} 
}

	 Defining the average density by \begin{equation}
	\bar{\rho} = {N \over L}
	\label{AA3}
	\end{equation}
	and the density $\rho(x) = n_i$ 
at a macroscopic coordinate $x={i \over L}$,
	one gets, in the large $L$ limit,   a linear steady state profile  $\langle \rho(x) \rangle  $ and long-range correlations   (for $y \neq x$)

		\begin{align}
	\langle \rho(x) \rangle &= 2 \bar{\rho} {G(1-x)   +H x \over G+H} \\
		\langle \rho(x) \rho(y) \rangle_c &=   {1 \over L}\left[ \langle \rho(x) \rangle\,  \delta(x-y)- {\langle \rho(x) \rangle \, \langle\rho(y) \rangle \over \bar{\rho}} \right]
	\label{AA4}
	\end{align}
	These long-range correlations are simply due to the fact that the total number $N$ of particles is fixed.

		\subsection{The distribution of the current for a single particle }

 To calculate the distribution of the  current in the steady state 
let us first consider the case of a single particle on a ring of $L$ sites. 
It is well known that for general Markov processes, the distribution of currents  (in the long time limit) can be determined by calculating the largest eigenvalue of a titled Markov matrix
\cite{DL,Touchette,Les Houches}.
In the case of a single particle on the ring with a battery,
 the long time asymptotics of the generating function of the integrated current $Q_t$ (here $Q_t$ is the number of jumps of the particle from site $L$ to  site $1$ minus the number of jumps from site $1$ to site $L$) is given by
		\begin{equation}
	\label{AA5}
	\lim_{t \to \infty} {1\over t } \log \langle e^{\lambda Q_t} \rangle = \nu(\lambda)
		\end{equation}
	where $\nu(\lambda)$  is the largest eigenvalue of the following $ L \times L$  tilted Markov matrix $M_\lambda$ 

		\[
		M_\lambda = 
			\left( {\begin{array}{cccccc}
					-H-1 & 1  &  0 &\cdots & 0 & G \, e^\lambda\\
					1 & -2& 1  & 0 &  \cdots  & 0\\
					0 & 1& - 2 & 1     &  \ddots  & 0\\
					\vdots & \ddots & \ddots & \ddots & \ddots & 0 \\
					0   & \ddots   &  0 &1 & -2 &1  \\
					H \, e^{-\lambda}  & 0  &  0 &\cdots & 1 &- G-1 
					\end{array} } \right)
			\]
Each   eigenvalue $\nu(\lambda)$  of $M_\lambda$ and the corresponding right 
			and left eigenvectors $r_i$ and $\ell_i$ (for $1 \le i \le L$) can be written  as
			\begin{align}
			\nu(\lambda) &= 2  \cosh \theta -2 \nonumber \\
			 r_i=  \cosh(\theta \, i + \varphi)  &		\quad \quad	\ell_i=  \cosh(\theta \,i + \psi)  \nonumber
			 \end{align}
where $\theta$ and $\varphi$ satisfy some coupled non-linear  equations
	\begin{align}
	 G e^\lambda  \cosh(L \theta+ \varphi) &= (H-1) \cosh(\theta + \varphi) + \cosh \varphi \nonumber  \\
	 H  e^{-\lambda}  \cosh( \theta+ \varphi) &= (G-1) \cosh(L \theta + \varphi)  \nonumber \\
	   & \quad + \cosh( (L+1)\theta + \varphi) \nonumber \\
	 H  e^{-\lambda}  \cosh( L \theta+ \psi) &= (H-1) \cosh( \theta + \psi) + \cosh \psi  \nonumber \\
	 G  e^{\lambda}  \cosh(  \theta+ \psi) &= (G-1) \cosh( L \theta + \psi) \nonumber \\
	 & \quad+ \cosh( (L+1) \theta + \psi) \nonumber 
	\end{align}
In the large $L$ limit, using the macroscopic  coordinate $x=i/L$  and $\theta= {\alpha /  L} $ one gets that 
			where $\alpha, \varphi$ and $\psi$ satisfy
\begin{align}
&			 G e^{\lambda} \cosh(  \alpha + \varphi) = H \cosh( \varphi)  \label{AA5bisa} \\
&			  e^{\lambda} \sinh(  \alpha + \varphi) =  \sinh( \varphi)    \label{AA5bisb} \\
&			    \cosh (  \alpha + \psi) =  e^\lambda \cosh( \psi) \label{AA5bisc}\\
&			 H   \sinh (  \alpha + \psi) =  G e^\lambda \sinh( \psi) \label{AA5bisd}
\end{align} 
leading to, for the largest eigenvalue of $M_\lambda$  and the corresponding eigenvectors
	\begin{align}
	 \nu(\lambda) & \simeq {\alpha^2 \over L^2}  \label{AA6a}  \\
		r_i\simeq  \cosh(\alpha x + \varphi) &  \quad \quad 
		\ell_i \simeq \cosh(\alpha x + \psi) 
		\label{AA6b} 
	\end{align}
	with
	\begin{align}
	\cosh \alpha  &= {G e^\lambda + H e^{-\lambda} \over G+H} \label{AA7a}\\
		\tanh  \, \varphi &= {H -  G \, e^\lambda \, \cosh \alpha \over G e^\lambda \, \sinh \alpha} \label{AA7b}\\
		\tanh \, \psi &= {e^\lambda  - \cosh \alpha \over \sinh \alpha}
	\label{AA7c}
	\end{align}
\ \\
{\bf Remark:} 
imagine now that   one weights all possible  trajectories of the particle by $e^{\lambda Q_t}$ where $Q_t$  is the integrated current through bond $(L,1)$ during the time interval $(0,t)$.
One can then try to determine    the probability 
that the particle is on site $i$, for these weighted trajectories at some time $\tau$.
When $ \tau \in (0,t) $ is such that  both  $\tau $, and $t-\tau$ are  very large, 
as for general Markov  processes  \cite{Touchette,Touchette-bis,Touchette-ter,DS},
  this probability is 
$$
p_i^{(\lambda)} =  { l_i \, r_i \over  \sum_j l_j \, r_j}
$$
and 
one can  check using Eqs.~(\ref{AA6a})-(\ref{AA7c}) that in the large $L$ limit
\begin{equation}
{p_L ^{(\lambda)}\over p_1^{(\lambda)}} \to  {\cosh(\alpha+\varphi) \, \cosh(\alpha + \psi) \over \cosh(\varphi) \, \cosh(\psi) } = {H \over G} 
\label{AA8}
\end{equation}

 In fact one can show using also Eqs.~(\ref{AA5bisa})-(\ref{AA5bisd}) that  $p_1^{(\lambda)} \to {2 G \over G+H}$ and $p_L^{(\lambda)}\to {2 H \over G+H}$ independent of $\lambda$.

\subsection{The case of $N$ non-interacting particles}
\label{2.3}
		As the particles are independent, one can immediately claim  that for a system of $N$ non-interacting particles 
		at density $ \bar{\rho} =N/L $, the generating
		 function of the integrated current $Q_t$ is given by
		\begin{equation}
	\lim_{t \to \infty} {1\over t } \log \langle e^{\lambda Q_t} \rangle = \mu(\lambda)= N \nu(\lambda)
		\label{AA9}
	\end{equation}
	so that
		\begin{equation}
		\mu(\lambda)  = {\bar{\rho} \over L} \left[ \cosh^{-1}\left(
				{G e^\lambda + H e^{-\lambda} \over G+H}\right) \right]^2
	       \end{equation}
		which can be rewritten as
		\begin{align}
	\mu(\lambda) &= { \bar{\rho} \over L} \left[\log \left( x + \sqrt{x^2-1}  \right) \right]^2  \label{AA10} \\ 
		\text{with} \ \ \ \ x &= {G e^\lambda + H e^{-\lambda} \over G+H} \nonumber 	
	\end{align}
	Note that he above formulas remain valid even when,
		for some choices of the parameters,
			$\cosh \alpha $ in Eqs.~(\ref{AA7a})-(\ref{AA7c}) is less than 1, in which case, $\alpha$ becomes imaginary or $x$ is less than 1.
\ \\			
				{\bf \noindent Remark:}  A consequence of (\ref{AA9}) is that, for large $L$,  all the cumulants of $Q_t$ scale as $1/L$. For example when $H=0$, i.e. when the bond $(L,1)$ is fully directed, one  gets, by expanding (\ref{AA10}) in powers of $\lambda$,
	that the first cumulants of the current are given,  in the long time limit, by
	\begin{align} 
	{\langle Q_t \rangle_c \over t} ={2 \bar{\rho} \over L}  
	\ \ \ & ; \ \ \
	{\langle Q_t^2 \rangle_c \over t} ={4 \bar{\rho}  \over L} \times  {1 \over 3} \label{cumulantsXa} \\
	{\langle Q_t^3 \rangle_c \over t} ={8 \bar{\rho}  \over L} \times {1 \over 15}
	\ \ \ &; \ \ \
	{\langle Q_t^4 \rangle_c \over t} ={16 \bar{\rho}  \over L} \times \left(-{1 \over 105} \right)
	\label{cumulantsXb}
	\end{align}

		\ \\ 
		{\bf Remark:} It is known  \cite{Bodineau1,Derrida-doucot-Roche} 
 that in the case of the SSEP  in contact  with  reservoirs at density $\rho_a$  at its left boundary and $\rho_b$ at its right boundary,
		\begin{equation}
		\mu(\lambda)= {1 \over L} \left[\log\left(\sqrt{1 + \omega     } + \sqrt{\omega     }\right)\right]^2
		\end{equation}
    with $\omega     =(z-1)(\rho_a z -\rho_b-\rho_a \rho_b (z-1))/z$.
			This  can be rewritten as
	\begin{equation} 
	\mu(\lambda) = {1 \over L} \left[ \log \left( x + \sqrt{x^2-1} \right) \right]^2 
	\label{AA11}
	\end{equation}
	with $x = \sqrt{ (\rho_a e ^\lambda +1 - \rho_a) (\rho_b e^{-\lambda} +1 - \rho_b)}$.
			There is a remarkable  similarity between this expression for the  SSEP with open boundary conditions and where particles have hard core interactions and (\ref{AA10}) where the particles are on a ring with a batttery and do not interact.
			
			\ \\ 
		 {\bf Remark:} 
		The reduction of the cumulants relative to the Poissonian statistics of the non-interacting open system can be understood from the constraint imposed by the 
		fixed number of particles on the ring, where transport involves repeated passages of the same particles. In the open system, particles are emitted by the electrode according to a Poisson process. Since they propagate independently, their travel-time fluctuations average out in the long-time counting statistics, and the cumulants of the transferred charge are determined by the Poissonian emission process. On the ring, by contrast, each particle ``enters'' the system again as soon as it ``exits'' through the unidirectional link. Its next passage therefore requires another complete traversal, so that successive traversal times add up. These repeated passages define a renewal process for each particle, yielding the cumulants given above.

		 While the constraint of a fixed number of particles on the ring is different from the usual exclusion present in the SSEP of the open system, in both cases 
		 the travel time of
		 one particle affects the arrival times of the following particles, which might explain the similarity between the generating functions of the two cases.
		 In Sec.~\ref{sec:MFT1}, we will show that the macroscopic fluctuations theory allows to get a more quantitative argument for this similarity 
		  (see the remark below Eqs.~(\ref{BB6})-(\ref{BB8})).
		  
		 \ \\   {\bf Remark:} for non-interacting particles, it is possible to extend the above calculations to the case of several directed bonds  which do not need to have all the same strength. In fact this  is a special case of the problem of a random walk on a  ring with arbitrary jumping rates \cite{derrida1983}.  
		For example, if there are $n$ equidistant directed bonds on a ring of length $L$, the generating function becomes $\mu_{n}$
		 with
		 \begin{equation}
		 \mu_{n}(\lambda) = n^2 \mu(\lambda/n) 
		 \end{equation}
         where $\mu(\lambda)$ is given by Eq.(\ref{AA10}). 
         In particular, the cumulants given by Eqs.~(\ref{cumulantsXa})-(\ref{cumulantsXb}) become
         \begin{align} 
	{\langle Q_t \rangle_c \over t} =n {2 \bar{\rho} \over L}  
	\ \ \ & ; \ \ \
	{\langle Q_t^2 \rangle_c \over t} ={4 \bar{\rho}  \over L} \times  {1 \over 3} \\
	{\langle Q_t^3 \rangle_c \over t} = \frac{1}{n}{8 \bar{\rho}  \over L} \times {1 \over 15}
	\ \ \ &; \ \ \
	{\langle Q_t^4 \rangle_c \over t} =\frac{1}{n^2}{16 \bar{\rho}  \over L}  \left(-{1 \over 105} \right)
	\label{cumulantsXn}
	\end{align}

		\ \\ 
		{\bf Remark:}  In the steady state, because the particles do not interact, the large deviation function ${\cF}$ of the density defined by
$$\text{Pro}(\{\rho(x)\}) \sim \exp[-\, L \,{\cF}_\text{steady state} \big(\{ \rho(x) \}\big) \, ]$$ 
is given by
$$ {\cF}_\text{steady state}  \big(\{ \rho(x) \}\big) = \int_0^1 dx \left[ 
1-{\rho(x) \over \langle \rho(x) \rangle}  
\log\left( {\rho(x) \over \langle \rho(x) \rangle} \right)  \right] $$
(note that the local form of  this large deviation function does not contradict the above expression of the long-range correlations (\ref{AA4}) which are simply due to the fact that the global density  $\bar{\rho} = \int_0 \rho(x) dx$ is fixed).

\ \\ \ \\

      \section{The macroscopic fluctuations theory for a ring with a battery}
      \label{sec:MFT1}

       In this section, we apply the macroscopic fluctuations theory (MFT) \cite{BDGJL1,BDGJL2,BDGJL6}
 to the case of a diffusive system of particles on a ring with a battery,
to calculate the fluctuations of the current~\cite{Bodineau1,BDGJL-2005-PRL,Bodineau-Derrida-2005,BDGJL-2006-JSP,Bodineau-Derrida-2007,Derrida 2007}. 
 In particular we obtain analytical expressions for  the average current along the ring and its fluctuations in different regimes.
     
	The macroscopic fluctuation theory (MFT) \cite{BDGJL1,BDGJL2,BDGJL6,Derrida 2007,Les Houches} has been widely used to compute the fluctuations and the large deviations  of the current of one dimensional diffusive systems in contact with two reservoirs 
\cite{Bodineau1,BDGJL-2005-PRL,BDGJL-2006-JSP,Bodineau-Derrida-2007,Derrida 2007},
  on a ring \cite{Appert,Bodineau-Derrida-2005},  on the infinite line  \cite{Antoine,Berlioz,Bet1,Krapivsky,Mallick} or   in higher dimensions \cite{Akkermans-Bodineau-Derrida-Shpielberg,Bodineau-2026}.

		In Appendix \ref{MFT} we argue, using the MFT,  that for a general diffusive system on a ring of size $L$ 
characterized by its diffusion constant $D(\rho)$ and its mobility $\sigma(\rho)$,
with a  battery located at the macrocospic position $x=0$  
 the generating function  of   the integrated current is given  for large $L$ by
\begin{equation}
\label{BB1}
 \lim_{t \to \infty}  {\log \langle e^{\lambda Q_t } \rangle  \over t} = \mu(\lambda)  
\end{equation}
where  $\mu(\lambda)$ can be obtained  by a variational expression  
\begin{equation}
\label{BB2}
\mu(\lambda)= {1 \over L} \max_{q, \{\rho(x)\}} \left[ \lambda q - \int_0^1 dx  {\Big(q+D\big(\rho(x)\big)\, \rho'(x) \Big)^2 \over 2 \sigma\big(\rho(x)\big)}\right]
\end{equation}
and the maximum is over all density profiles $\rho(x)$ which satisfy the following two constraints.
\begin{equation}
\label{BB3}
\bar{\rho} = \int_0^1 \rho(x) dx \ \ \ \ ; \ \ \ \  B=\int_{\rho(1)}^{\rho(0)} {2 D (r) \over \sigma(r)}
dr   \ . 
\end{equation}
The first constraint is simply due to the fact that the number of particles $N= L \bar{\rho}$  on the ring is fixed   while the second constraint is a characteristic of the  battery (one can think of $B$ as being its  strength).

As $\lambda$ varies, the current $q$, the optimal profile $\rho(x)$ as well as $\rho(0)$ and $\rho(1)$ in (\ref{BB2}) vary  whereas the global density $\bar{\rho}$ and  the strength $B$ of the battery  remain fixed.

One way of finding the optimal profile $\rho(x)$ in (\ref{BB3}) is, first,  to minimize   the large deviation function $I(q)$ of the integrated current
with the constraint (\ref{BB3})
\begin{equation}
I(q)=\min_{\{\rho(x)\}}\int_0^1 dx  {\Big(q+D\big(\rho(x)\big) \,\rho'(x) \Big)^2 \over 2 \sigma\big(\rho(x)\big)} 
\label{BB5}
\end{equation}  and then to optimize (\ref{BB2}) over $q$.
It is easy to verify from (\ref{BB5}) that $I(q)$ satisfies the fluctuation theorem 
\begin{equation}
\label{BB4}
I(q)=I(-q) + q \int_0^1 {2D\big(\rho(x)\big)  \over \sigma\big(\rho(x)\big)} \rho'(x) dx  = I(-q)-B  \, q 
\end{equation}
(because the  optimal profile $\rho(x)$ in (\ref{BB2}) is unchanged  when $q $ is replaced by $-q$)
and this implies that $\mu(\lambda)=\mu(-\lambda-B) $ which is another way of writing the fluctuation theorem.

Because the  optimisation  of of $I(q)$, which is  a functional of the density  $\rho(x)$ and of its derivative $\rho'(x)$, is a  problem  very similar to the optimisation of a Lagrangian in classical mechanics, one can show that
the optimal profile satisfies
$${(D(\rho(x)) \rho'(x))^2 \over 2 \sigma(\rho(x))} - {q^2 \over 2 \sigma(\rho(x))} - C_1 \rho(x) =C $$
where $C_1$ is a Lagrange multiplier associated to the constraint (\ref{BB3}) that the total density $\bar{\rho} $ is fixed.
Assuming that $B >0$ and that the optimal profile is monotone, one can then obtain $I(q)$ in a parametric form by varying the parameters $K$ and $K_1$ (where $K=2C/q^2$ and $K_1=2 C_1/q^2)$
\begin{align}
q&= \int_{\rho(1)}^{\rho(0)}  {1 \over  \sqrt{1 + (K+ K_1 \rho) \sigma(\rho)}}  D(\rho) \, d\rho
 \label{BB6}\\
I(q)&= q \Big[-{B \over 2} \nonumber \\
  &  + \int_{\rho(1)}^{\rho(0)} { 2 + (K+K_1 \rho) \sigma(\rho) 
\over  2 \sigma(\rho) \sqrt{1 + (K + K_1 \rho) \sigma(\rho)}} 
D(\rho) d \rho  
\Big] 
\label{BB7} 
\\ \bar{\rho}&  = {1 \over q} \
 \int_{\rho(1)}^{\rho(0)}  { \rho \over    \sqrt{1 + (K+ K_1 \rho) \sigma(\rho)}}   D(\rho) d \rho
\label{BB8} 
\end{align}
So to obtain $I(q)$ given by the Eq.~(\ref{BB7}) one needs to find, for each choice of $\rho(0)$ and $\rho(1)$, 
 the constants $K$ and $K_1$ 
which give in (\ref{BB6},\ref{BB8}) 
 the desired values of $q$ and $\bar{\rho}$ 
and then to optimize over $\rho(0)$ and $\rho(1)$ 
under the constraint (\ref{BB3}).
\ \\ \ \\
{\bf Remark:} One can notice the similarity between (\ref{BB7},\ref{BB8}) with the corresponding  expressions (Equations (17,18) of Ref. \onlinecite{Bodineau1}) for the system with open boundaries. The main differences  are that here one has to optimize over the densities $\rho(0)$ and $\rho(1)$ keeping  the constraints (\ref{BB3}) and that there is an extra constant $K_1$ which allows  to keep the total density $\bar{\rho}$ fixed. One can argue that the reason why (\ref{AA10}) obtained for $N$ non-interacting particles with a battery  was very similar to the corresponding quantity for the SSEP with open boundaries is that due to  the constant $K_1$,
the argument of the square root in (\ref{BB6},\ref{BB7},\ref{BB8})  becomes quadratic in $\rho$ while  for the SSEP (see (\ref{ZZa},\ref{ZZb})) with open boundaries  it is also quadratic in  $\rho$ . 
\ \\ \ \\
The steady state current $q_0$ and steady state profile $\rho_0(x)$  are  given by 
 \begin{equation}
q_0= \int_{r_1}^{r_0} D(r) \, dr  
 \ \ \ \ \ ; \ \ \ \ \ 
D(\rho_0(x)) \, \rho_0'(x)=-q_0 
\label{BB9}
 \end{equation}
where $r_0 \equiv \rho_0(0)$ and $r_1 \equiv\rho_0(1)$ are solutions (see (\ref{BB3})) of
\begin{equation}
\label{BB10}
\bar{\rho} = 
{\int_{r_1}^{r_0} r \, D(r) \, dr  
\over \int_{r_1}^{r_0} D(r) \, dr  }
\ \ \ \ \text{and}  \ \ \ \  B=\int_{r_1}^{r_0} {2 D (r) \over \sigma(r)}
dr   \ .
\end{equation}

In the steady state, $I(q_0)=0$. 
For general $D(\rho)$ and $\sigma(\rho)$ the parametric 
form (\ref{BB6}-\ref{BB8}) is not very convenient to obtain  more explicit expressions for $I(q)$. However  one can expand  $I(q)$ around its minimum  at $q=q_0$. 
For example, expanding $I(q)$ 
(see Appendix \ref{derivation})
 up to order  $(q-q_0)^2$, one gets
\begin{widetext}
\begin{equation}
I(q) =
 {\big(I_1 (\sigma_0 - \sigma_1)   - I_0 (   r_0 \, \sigma_0 -    r_1 \, \sigma_1)\big)^2  
\, (q-q_0)^2 
\over  2 I_0 \big[
\big( J_0 (   r_0 \sigma_0
-     r_1 \sigma_1)
- 2 J_1 (\sigma_0  - \sigma_1) \big)
(   r_0 \sigma_0 -    r_1 \sigma_1) 
+J_2 (\sigma_0 - \sigma_1)^2  
\big]} 
\label{BB11}
\end{equation}
\end{widetext}
where 
 $ \sigma_0= \sigma(r_0)$, $\sigma_1=\sigma(r_1)$
and for all $p$ 
\begin{equation}
I_p = \int_{r_1}^{r_0} D(r) \, r^p \, d r  ,\quad
J_p = \int_{r_1}^{r_0} D(r)\, \sigma(r)  \, r^p \, d r  
\label{BB12}
\end{equation}
}

In the case of non-interacting particles
one has 
 $D(\rho)=1$ and $\sigma(\rho)=2 \rho$
so that 
$$r_0= {2 \bar{\rho} e^B \over e^B +1} 
 \ \ \ \ \ ; \ \ \ \ \ r_1= {2 \bar{\rho}  \over e^B +1} $$
and 
\begin{align}
\lim_{t\to \infty} {\langle Q_t \rangle \over t} &= { 2 \bar{\rho}\over L} \  {e^B-1 \over e^B+1}  \nonumber \\
\lim_{t\to \infty} {\langle Q_t^2 \rangle_c \over t} &= {4  \bar{\rho} \over 3 L } {e^{2 B} + 4 e^B + 1 \over (e^{B}+1)^2}  \nonumber
\end{align}
This  agrees with Eq.(\ref{AA10}) using the fact that 
 $ e^B={G \over H} = {r_0\over r_1}$ (which was already noticed in Eq.(\ref{AA8})).

\section{The symmetric exclusion process with a battery}
\label{sec:SSEP}
 
In this section, we consider the symmetric simple exclusion process (SSEP)
\cite{KOV,Goncalves,De Masi-Presutti-Tsagkarogiannis-Vares}
  on a ring with one special link playing the role of a battery.
The dynamics are the same as in the case of non-interacting particles discussed in section \ref{sec:noninteracting} except that, in the SSEP, two particles cannot occupy the same site.
 We use Monte-Carlo numerical simulations to obtain results for rings of finite sizes up to $L=512$,
       which are compared with the analytical predictions of MFT. 
The effect of the battery is that a particle on site $L$ jumps to site 1 at rate $G$ and a particle on site $1$ jumps to site $L$ at rate $H$.
Following the assumptions  (\ref{ZZ6}) of the Appendix,  the optimal profile in (\ref{ZZ5}) should satisfy
\begin{equation}
G\,  \rho(1)\, (1-\rho(0)) = H \, \rho(0)\,  (1- \rho(1))
\label{DD0}
\end{equation}

Using the fact that $D(\rho)=1$ and $\sigma(\rho)=2 \rho(1-\rho)$ for the SSEP,  one has $G/H = \log B$. Moreover (\ref{BB2},\ref{BB3}) and (\ref{BB6},\ref{BB7},\ref{BB8}) give in principle $I(q)$ and $\mu(\lambda)$ in a parametric form.
Even for the SSEP we did not succeed to obtain  more explicit expressions of these quantities for arbitrary $q$ or $\lambda$. However as explained in section \ref{sec:MFT1}  one can   obtain the first cumulants of the current in several cases.   

\subsection{  The half-filled case} \hfill\\
For 
 \begin{equation}
 \bar{\rho}={1 \over 2}  \ \ \ \ \ \text{and} \ \ \ \ \ G=1-H 
 \end{equation}
one has $\rho(1)=1-\rho(0)$ 
for symmetry reasons, 
and one gets from (\ref{BB9},\ref{BB11})
\begin{align}
\label{DD1} {\langle Q_t\rangle  \over t } &= 2 \rho(0)-1   ; \quad  
{\langle Q_t^2 \rangle_c\over t }=  {1 + 2 \rho(0)  -2  \rho(0)^2 \over 3 }  \\
 &  \text{where}  \ \ \ \rho(0) =
{  \sqrt{G}  \over 
 \sqrt{G} + \sqrt{1-G}} 
\end{align}
\begin{figure*}[h!]
    \centering
    \includegraphics[width=.66\textwidth]{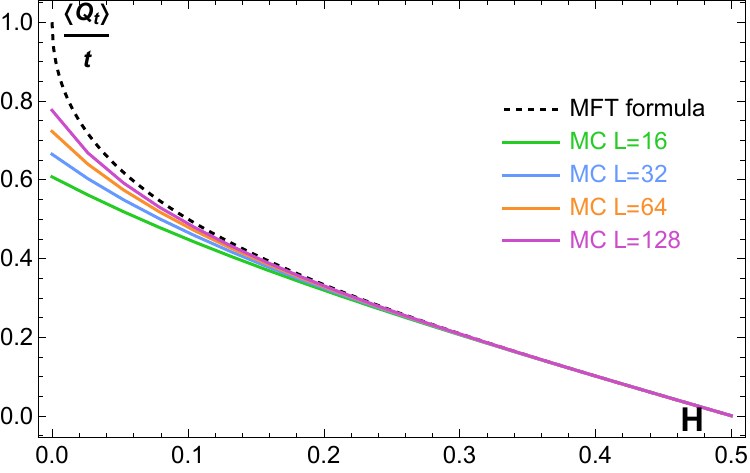}
    \caption{
 Average current  (multiplied by $L$) as a function of $H=1-G$, for the SSEP on a ring at density $\bar{\rho}=1/2,$ 
for system sizes $L=16,32,64,128$ (obtained by Monte Carlo, with increasing $L$ corresponding to higher curves). The dashed line is the prediction (\ref{DD1})  of the MFT.}
    \label{Fig1}
\end{figure*}
\begin{figure*}[h!]
    \centering
    \includegraphics[width=.66\textwidth]{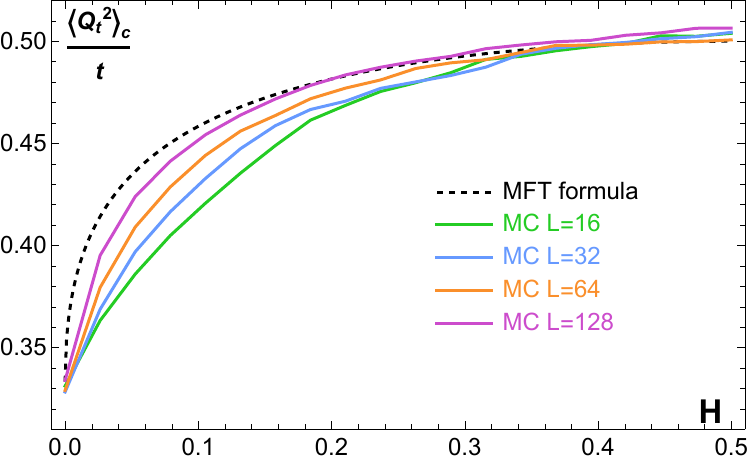}
    \caption{
 Second cumulant of the current  (multiplied by $L$) as a function of $H=1-G$ 
for the SSEP on a ring at density $\bar{\rho}=1/2$,
for system sizes $L=16,32,64,128$ (obtained by Monte Carlo, with increasing \(L\) corresponding to higher curves). The dashed line is the prediction (\ref{DD1}) of the MFT.}
    \label{Fig2}
\end{figure*}
Figures \ref{Fig1} and \ref{Fig2} show the first two cumulants of the current obtained by Monte Carlo simulations versus $H$ in the half-filled case for system sizes up to $L=128$. The results seem to converge to the predictions of the MFT. However the convergence looks much slower when $H$ is small, i.e. when the effect of the battery is very strong.

\subsection{ The fully directed battery}  
\begin{figure*}[h!]
    \centering
    \includegraphics[width=.66\textwidth]{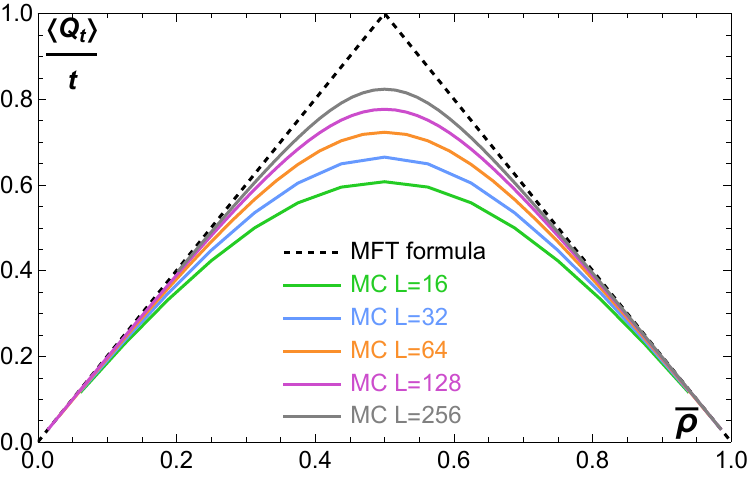}
    \caption{
Average current  (multiplied by $L$) as a function of   the density $\bar{\rho}$ for the SSEP on a ring  with a fully directed battery ($G=1, H=0$),
for systems sizes  $L=16,32,64,128,256$ (obtained by Monte Carlo, with increasing $L$ corresponding to higher curves). The dashed line is  the prediction (\ref{DD2a})-(\ref{DD2b})  of the MFT.
}
    \label{Fig3}
\end{figure*}

\begin{figure*}[h!]
    \centering
    \includegraphics[width=.66\textwidth]{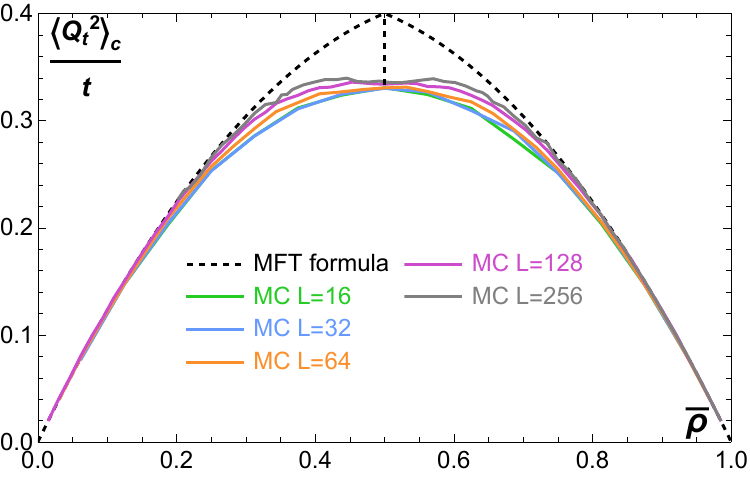}  
    \caption{
Second cumulant of the current  (multiplied by $L$) as a function of the density $\bar{\rho}$ for the SSEP on a ring  with a fully directed battery ($G=1, H=0$),
for systems sizes  $L=16,32,64,128,256$ (obtained by Monte Carlo, with increasing $L$ corresponding to higher curves). The dashed line is  the prediction (\ref{DD2a})-(\ref{DD2b})  of the MFT, including the special value 1/3 at $\bar{\rho}=1/2$, as predicted by Eq.(\ref{DD1}).} 
    \label{Fig4}
\end{figure*}

\hfill \\
For the SSEP, when the bond describing the effect of battery 
is fully directed, i.e. when
$$G=1 \ \ \ \ \ ; \ \ \ \ \ H=0$$
the MFT prediction  of section \ref{sec:MFT1} 
leads to
\begin{align}
\label{DD2a}
&\bar{\rho} \le {1 \over 2} : & {\langle Q_t\rangle  \over t }
& = 2 \bar{\rho}&   ;
& {\langle Q_t^2 \rangle_c\over t }
 =  {20 \bar{\rho}- 16 \bar{\rho}^2 \over 15} \\
\label{DD2b}
&\bar{\rho} \ge {1 \over 2}  : & {\langle Q_t\rangle  \over t }
 & = 2 (1-\bar{\rho})   &  ;  
& {\langle Q_t^2 \rangle_c\over t }
 =  {4+12 \bar{\rho}- 16 \bar{\rho}^2 \over 15 }   
\end{align}
 Figures~\ref{Fig3} and \ref{Fig4}  show the first two cumulants of the current obtained by Monte Carlo simulations versus $\bar{\rho}$ in the fully directed case for system sizes up to $L=256$.   The results seem to converge to the predictions of the MFT, with a much slower convergence near half-filling ($\bar{\rho}=1/2$).

\newpage
\subsection{  Behavior near half-filling} \hfill \\
Comparing Eq.~(\ref{DD1}) and Eqs.(\ref{DD2a})-(\ref{DD2b}) one gets two conflicting  predictions for ${\langle Q_t^2 \rangle_c \over t}$
when $\bar{\rho}=1/2$ and $G=1$  both coming from the general expression (\ref{BB11}) : replacing $G$ by $1$ in Eq.~(\ref{DD1}) gives $1/3$ while replacing $\bar{\rho}$ by $1/2$ gives $2/5$. This discrepancy comes from the fact that the limits $\bar{\rho} \to 1/2$ and $G\to 1$ do not commute. In fact when $G$ is close to $1$ and $\bar{\rho}$ is close to $1/2$, one can show using Eq.~(\ref{BB11}) that 
\begin{equation}
{\langle Q_t^2 \rangle_c\over t }=  {5(1-G) + 6 (\bar{\rho}-{1 \over 2})^2  \over
15[ (1-G) +  (\bar{\rho}-{1 \over 2})^2 ]}
\label{DD3}
\end{equation}

This would imply that, for $G=1$,  there is an extra point in the prediction  of figure  \ref{Fig4} with the value $1/3$ at $\bar{\rho}=1/2$ illustrating the fact that the second cumulant has a limit $2/5$ as $\bar{\rho} \to 1/2$ but takes the value $1/3$ when $\bar{\rho}=1/2$.

\begin{figure*}
\centerline{ \includegraphics[width=0.66 \textwidth]{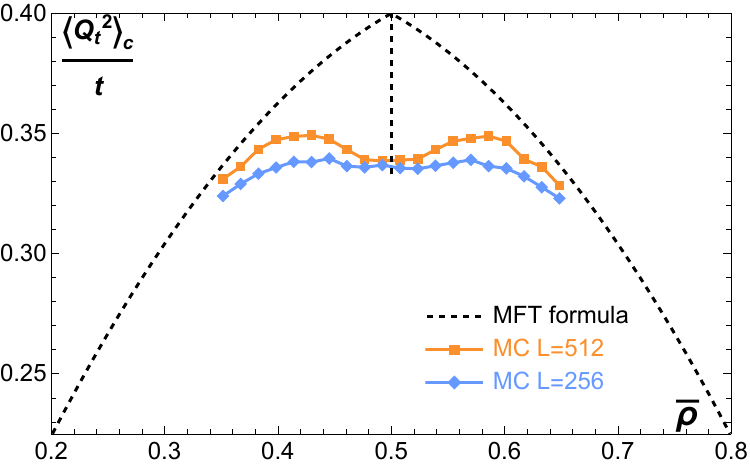} }
\caption{ Second cumulant of the current  (multiplied by $L$) as a function of  the density $\bar{\rho}$ for the SSEP on a ring 
 with a fully directed battery ($G=1, H=0$),
for systems sizes  $L=256$ and $512$ (obtained by Monte Carlo), for $0.35 < \bar{\rho} < 0.65$.
For $L=512$, one clearly sees that the behavior is non-monotonous, with the second cumulant larger than the value 1/3 for $\bar{\rho}$ close to 0.4-0.45, and going down to the value $\sim 1/3$ for $\bar{\rho}$ close to 1/2.
The dashed line is  the prediction (\ref{DD2a})-(\ref{DD2b})  of the MFT, including the special value 1/3 at $\bar{\rho}=1/2$, as predicted by Eq.(\ref{DD1}).}
\label{fig:C2L512}
\end{figure*}

For finite-size rings, this implies that the second cumulant is non-monotonous as a function of $\bar{\rho}$ for $0<\bar{\rho}<1/2$, with the value $\sim1/3$ close to $\bar{\rho}=1/2$.
In order to check this numerically, we had to consider relatively large rings ($L \simeq 500$), requiring a larger numerical effort. Fig.\ref{fig:C2L512} shows the second cumulant as a function of $\bar{\rho}$ for $L=256$ and $L=512$, for 
$0.35<\bar{\rho}<0.65$.  
One can see that the second cumulant is larger than 1/3 for $\bar{\rho}$ close to 0.4-0.45, and goes down to $\simeq 1/3$ for $\bar{\rho}$ close to 1/2.
This effect is small (but visible) for $L=256$, and clearly visible for $L=512$.
 From these data, one can expect that for rings of larger size, the second cumulant will reach higher values, and will decrease more abruptly
to the value $1/3$ for $\bar{\rho}$ close to $1/2$, in a interval which becomes narrower as $L$ increased. Ultimately, according to our
theoretical predictions based on the MFT, for $L \to \infty$ the second cumulant should
follow Eqs.(\ref{DD2a})-(\ref{DD2b}), except at $\bar{\rho}=1/2$ exactly where it will have the value 1/3.

\subsection{ Convergence of the results}
\begin{figure}[h!]
    \centering
    \includegraphics[width=.45\textwidth]{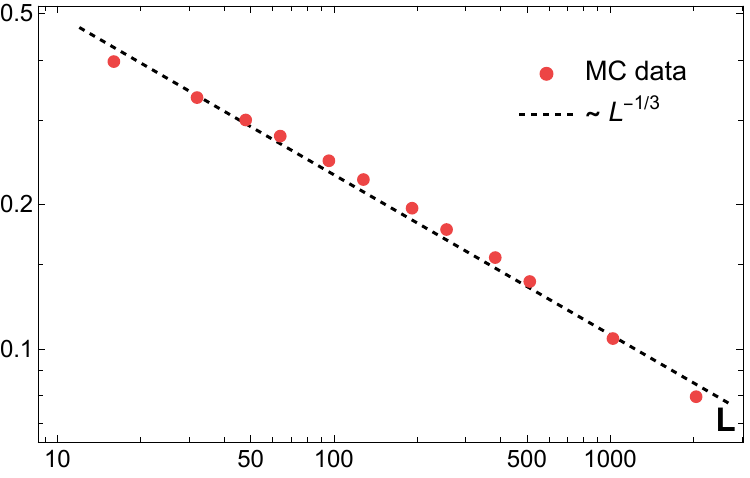}
    \caption{
Log-Log plot of the  difference in Eq.(\ref{DD4})  as a function of the system size $L= 16,32,64,\cdots 2048$, for the SSEP on a ring  with a fully directed battery ($G=1, H=0$)  and at density $\bar{\rho}=1/2$.  Despite a slight curvature in this Log-Log plot, the data seem consistent with
a power law  convergence $\sim L^{-1/3}$ as indicated in (\ref{DD4}).}
\label{Fig5}
\end{figure}

\hfill \\
Already for the average current, the convergence to the predictions looks very slow when $\bar{\rho}$  is close to $1/2$ or/and $G$ is close to 1.
Figure \ref{Fig5} shows a log-log plot of the difference between the results of the Monte Carlo simulations and the prediction that the average current is equal to 1 for $G=1$ and $\bar{\rho}=1/2$. This log-log plot indicates a power law convergence with an exponent close to $1/3$
\begin{equation}
L \left(\ \left. {\langle Q_t\rangle \over t} \right|_\text{theory}
-\left. {\langle Q_t\rangle \over t} \right|_\text{simulations} \ \right) \sim L^{-{1\over 3}}
\label{DD4}
\end{equation}
but we do not have at the moment any explanation for this apparent power law convergence.

\section{Conclusion}
\label{sec:conclusion}

In the present work we have shown
in section 3,  using the macroscopic fluctuation theory,
how to compute  (\ref{BB6},\ref{BB7},\ref{BB8})
  the large deviation  function of the current of a diffusive system on a ring with a battery at the origin.  Our computation is based on two assumptions (\ref{ZZ5},\ref{ZZ6}) discussed in the appendix \ref{MFT}, that the optimal profile is time independent (\ref{ZZ5}) and that the  strength of the battery determines the relation between the  densities of the optimal profile on the two sides of the battery.

For non-interacting particles, we checked that these  predictions agree with the results of the  direct exact calculation of section 2. To our surprise,  we noticed that the large deviation function of the current for non-interacting particles with a battery has a similar form as for the SSEP with open boundary conditions (see Eqs.(\ref{AA10}) and (\ref{AA11}) and the remark after Eq.(\ref{BB8})).

In section 4 we have compared the predictions of section 3 based on MFT to the results of Monte Carlo simulations  in the case of the SSEP. The agreement  looks satisfactory except when the effect of the  battery becomes very strong. In fact when the directed bond which represents the effect of the battery is fully directed (i.e. when $G=1$ and $H=0$), it is hard to tell  whether our simulations, for the system sizes we could achieve, will converge to the predictions of the MFT, and if they do, the convergence seems to be a power law which seems to depend on the global density $\bar{\rho}$ and is close to $1/3$  when $\bar{\rho}=1/2$.
At the moment we have no explanation for this possible power law.
Also, for the system sizes in our simulations, we were not able to observe  directly the scaling function (\ref{DD3}),
but we could see some non-monotonous behavior of the second cumulant as a function of the density close to half-filling (Fig.~\ref{fig:C2L512}).

It would be interesting to give a mathematical justification of the two assumptions  at the basis of our theoretical predictions. Also, one could try to test our predictions  on other  diffusive systems.

We have to admit that  the results of our Monte Carlo
simulations are less satisfactory than we expected. The reason is that we found it very difficult to obtain precise numerical values of the second cumulant of the current, even when we performed of order $10^{10}$-$10^{11}$ updates per site. We think that the reason is that it takes of order $L^2$ time steps to equilibrate the system, and many realizations are needed to obtain a good estimate of this variance. It would be interesting to see whether alternative ways of making these simulations would allow to explore accurately larger system sizes.

Finally, one can wonder whether the MFT developed here could be adapted to other situations such as the case of a  diffusive system with an active 
tracer.\cite{Miron,Miron2}

\begin{acknowledgements}{T.J. thanks N. Cl\'ement for the initial discussions that led to this project. }
\end{acknowledgements}

 \newpage
\appendix
\section{Macroscopic fluctuation theory}
\label{MFT}
 
For general diffusive  systems,  the large deviations function $I(q)$ of the current is given by a time and space dependent  variational problem
\cite{BDGJL-2005-PRL,BDGJL-2006-JSP,Derrida 2007, Les Houches,Bodineau-Derrida-2005}
involving only macroscopic observables and  two transport coefficients: the diffusion $D(\rho)$ and its mobility $\sigma(\rho)$, both indexed by the local density $\rho$.

In presence of a weak field $E(x)$ which varies slowly in space (i.e. which is a function of the macroscopic coordinate $x$)
it is of the form 
\begin{equation}
I(q) = \lim_{T \to \infty}  I_T(q)
\label{ZZc}
\end{equation}
with
\begin{widetext}
\begin{equation} 
I_T(q)= { 1 \over T}  \min_{\{\rho(x,\tau),j(x,\tau) \}}
\left[ \int_0^T  d \tau  \int_0^1 dx {\Big(j(x,\tau) +
D\big(\rho(x,\tau)\big)\ \rho'(x,\tau)  
- E(x) 
\sigma\big(\rho(x,\tau)\big) \Big)^2 \over 2 
\sigma\big(\rho(x,\tau)\big)  }   \right] 
\label{ZZ1} \end{equation}
\end{widetext}
 where the macroscopic current $j(x,\tau)$ and density $\rho(x,\tau)$ satisfy the continuity equation
\begin{equation} 
{d \rho\over d\tau} = - {d j \over dx}
\label{ZZ2} \end{equation}
and the constraint on the integrated current
\begin{equation} 
q={1 \over T} \int_0^T d \tau \int_0^1 j(x,\tau)
\label{ZZ3} \end{equation}
(one  way of justifying (\ref{ZZ1}) is to realize that it is consistent with
 the following hydrodynamics equations
$$j = - D(\rho) \rho' +  E(x) \sigma(\rho)+  \sqrt{ \sigma(\rho)} \, \eta(x,\tau)$$ where $\eta(x,\tau)$ is a white noise.)

\ \\ 
For {\it non-interacting  particles}
\begin{equation}
\label{ZZa}
D(\rho)=1 \ \ \ \ \ ; \ \ \ \ \ \sigma(\rho)=2 \rho
\end{equation}
 while for the SSEP
\begin{equation}
\label{ZZb}
D(\rho)=1 \ \ \ \ \ ; \ \ \ \ \ \sigma(\rho)=2 \rho (1-\rho)
\end{equation}

One can check that  the symmetry 
$\{q,j(x,\tau),\rho(x,\tau) \} \to \{ -q,-j(x,T-\tau),\rho(x,T-\tau)\} $ and  an integration by parts based on (\ref{ZZ2}) imply 
\begin{align}
I_T(q)=& I_T (-q)-2 q \int_0^1 E(x) dx  \nonumber \\ &+ {1 \over T} \int_0^1 dx [f( \rho(x,T) ) - f(\rho(x,0))]
\end{align}
 where $f''(\rho)={2 D(\rho) \over \sigma(\rho)} $.
Taking the $T \to \infty$ limit, one gets the following  form of the fluctuation theorem
\begin{equation}
 I(q) = I(-q) -2 q \int_0^1  E(x) dx 
\label{ZZ4}
\end{equation}

To  obtain the variational problem of section \ref{sec:MFT1} we will make two assumptions.
First, in the long time limit we assume that the optimal profile becomes time-independent. This implies that the variational  problem (\ref{ZZc},\ref{ZZ1}) becomes
\begin{equation}
I(q)=\min_{\{\rho(x)\}}\int_0^1 dx  {\Big(q+D\big(\rho(x)\big)\, \rho'(x)-  E(x) \sigma\big( \rho(x)\big) \Big)^2 \over 2 \sigma\big(\rho(x)\big)}
\label{ZZ5}
\end{equation}
This assumption, called the additivity principle \cite{Bodineau1,Hurtado-Garrido-PRL-(2009),Hurtado-Garrido-PRE-(2010),giardina,Saito-Dhar}
 was first introduced in the case of open systems in contact with two reservoirs. It is known to hold under some conditions on $D(\rho)$ and $\sigma(\rho)$  satisfied  for example by the SSEP, but phase transitions to regimes where the optimal profile becomes time-dependent have been predicted and observed numerically.
\cite{BDGJL-2005-PRL,BDGJL-2006-JSP,Bodineau-Derrida-2005,Baek1,Baek2,Hurtado-Garrido-PRL-(2011),Shpielberg-Yaroslav-Akkermans}

Second, we also assume that the case of a battery \big(where the asymmetry is limited to a single bond $(L,1)$\big)  can be obtained as the limiting case ($\epsilon \to 0$) of a localized field~\cite{battery}
$$E(x)={B \over 2 \epsilon} \ \ \ \ \ \text{for } \ \ \ \  x\in (0,\epsilon) \cup (1-\epsilon, 1)$$
and $E(x)=0$ \ for $ x\in(\epsilon,1-\epsilon)$.
For fixed $q$ and small $\epsilon$, the optimal profile $\rho(x)$ in (\ref{ZZ5})  should satisfy,    in the region where $E(x)$ is  non-zero,
$$D\big(\rho(x)\big) \rho'(x) = {B \over 2 \,  \epsilon } \sigma\big(\rho(x)\big)
$$
which gives  in the limit $\epsilon \to 0$
\begin{equation}
\int_{\rho(1-\epsilon)}^{\rho(\epsilon)} {2 D(\rho) \over \sigma(\rho)} d \rho = B
\label{ZZ6}
\end{equation}

\section{Derivation of the Eq.(\ref{BB11})}
\label{derivation}
The steady state  current $q_0$ and profile $\rho_0(x)$   satisfy  (\ref{BB9}) where $r_0\equiv \rho_0(0)$ and $r_1 \equiv \rho_0(1)$ are solutions of (\ref{BB10}), 
and one has $I(q_0)=0$ in (\ref{BB7}).
\ \\ \ \\

To obtain (\ref{BB11}) we assume that $\delta q=q-q_0$ is small and that $K,K_1, \delta \rho(x) \equiv \rho(x)-\rho_0(x)$ are  all of order $\delta q$.
Then (\ref{BB6}) tells us that

\begin{equation}
\delta q= D(r_0) \, \delta\rho(0) - D(r_1) \, \delta\rho(1)  -{K J_0 + K_1 J_1  \over 2}
\label{YY3}
\end{equation}
and (\ref{BB8}) that
\begin{equation}
{ D(r_0) \, r_0\,  \delta\rho(0)
- D(r_1) \, r_1 \,\delta\rho(1)
 \over I_0} - {I_1  \over I_0^2} \delta q -{ K J_1 + K_1 J_2  \over 2 I_0} =0
\label{YY4}
\end{equation}
where  as in (\ref{BB12}) 
\begin{equation}
I_p= \int_{r_1}^{r_0} \, r^p \, D(r) \, dr
\ \ \ \ \ \text{and} \ \ \ \ \ J_p= \int_{r_1}^{r_0} \, r^p \,D(r) \, dr
\label{YY5}
\end{equation}
Moreover the constraint due to the battery (see (\ref{BB3},\ref{ZZ2}))becomes 
\begin{equation}
{2D(r_0) \over \sigma(r_0)} \, \delta \rho(0)
={2D(r_1) \over \sigma(r_1)} \, \delta \rho(1)
\label{YY6}
\end{equation}
so that one can determine  $K,K_1$ and $\delta \rho(1)$ in terms of $\delta \rho(0)$ and $\delta q$. Then at order $\delta q^2$,  equation (\ref{BB7}) becomes
\begin{equation}
\label{YY7}
I(q)={q_0 \over 8} \int_{r_1}^{r_0} (K +K_1 r)^2 \, \sigma(r) \,  D(r) \, dr
\end{equation}
(which in fact  is a function of $\delta q$ and $\delta \rho(0)$). Then minimizing this quadratic form over $\delta \rho(0)$ leads to~(\ref{BB11}).

\section{Details on the numerical Monte-Carlo implementation}
We estimate the current and its fluctuations using the Gillespie algorithm \cite{Gillespie1977}, an exact continuous-time Monte Carlo method for Markov jump processes.
Starting from a configuration with a fixed number of particles, all allowed jumps are assigned their corresponding rates. 
At each step of the dynamics, the total escape rate \(R\) is computed, the time increment is drawn from an exponential distribution of mean \(1/R\), 
and one of the allowed jumps is selected with probability proportional to its rate. The configuration is then updated and the list of allowed transitions is modified accordingly. After an initial relaxation time $\tau_{\rm in}$, the integrated current \(Q_T\) through a fixed bond is recorded during a measurement time \(T_{\rm meas}\). Repeating this procedure over many independent trajectories $n_{sim}$ gives the mean current and the second cumulant as
\[
J = \frac{\langle Q_T\rangle}{T_{\rm meas}},
\qquad
C_2 = \frac{\langle Q_T^2\rangle-\langle Q_T\rangle^2}{T_{\rm meas}}.
\]
Ideally, the measurement time should be chosen large enough for finite-time
corrections to be negligible, while the number of independent trajectories
controls the remaining statistical uncertainty. In practice, for the data shown
in Figs.~2 and 3, we used $T_{\rm meas}=512\,{\rm k}$,
$\tau_{\rm in}=40\,{\rm k}$, and $n_{\rm sim}=250\,{\rm k}$ for the
largest system sizes. For Figs.~4 and 5, we used
$T_{\rm meas}=80\,{\rm k}$, $\tau_{\rm in}=20\,{\rm k}$, and
$n_{\rm sim}=320\,{\rm k}$ for $L\leq 128$. For $L=256$, we increased
the measurement time to $T_{\rm meas}=320\,{\rm k}$, with
$\tau_{\rm in}=80\,{\rm k}$ and $n_{\rm sim}=80\,{\rm k}$. Finally,
Fig.~6 required the largest numerical effort: for $L=512$, we used
$T_{\rm meas}=640\,{\rm k}$, $\tau_{\rm in}=160\,{\rm k}$, and
$n_{\rm sim}=160\,{\rm k}$. In all cases, we checked that further increasing
$T_{\rm meas}$ did not change the results significantly within statistical uncertainties.


\begin{thebibliography}{00}


\bibitem{BDGJL2}
Bertini, L., De Sole, A., Gabrielli, D., Jona-Lasinio, G., Landim, C.,
\textit{Macroscopic fluctuation theory for stationary non-equilibrium states},
Journal of Statistical Physics \textbf{107}, 635--675 (2002).

\bibitem{BDGJL6}
Bertini, L., De Sole, A., Gabrielli, D., Jona-Lasinio, G., Landim, C.,
\textit{Macroscopic fluctuation theory},
Reviews of Modern Physics \textbf{87}, 593--636 (2015).

\bibitem{Bodineau1}
Bodineau, T., Derrida, B.,
\textit{Current fluctuations in nonequilibrium diffusive systems: An additivity principle},
Physical Review Letters \textbf{92}, 180601 (2004).

\bibitem{Derrida 2007}
Derrida, B.,
\textit{Non-equilibrium steady states: Fluctuations and large deviations of the density and of the current},
Journal of Statistical Mechanics: Theory and Experiment \textbf{2007}, P07023 (2007).

\bibitem{battery}
Bodineau, T., Derrida, B., Lebowitz, J.~L.,
\textit{A diffusive system driven by a battery or by a smoothly varying field},
Journal of Statistical Physics \textbf{140}, 648--675 (2010).

\bibitem{Sadhu}
Sadhu, T., Majumdar, S.~N., Mukamel, D.,
\textit{Long-range steady-state density profiles induced by localized drive},
Physical Review E \textbf{84}, 051136 (2011).

\bibitem{BDGJL1}
Bertini, L., De Sole, A., Gabrielli, D., Jona-Lasinio, G., Landim, C.,
\textit{Fluctuations in stationary nonequilibrium states of irreversible processes},
Physical Review Letters \textbf{87}, 040601 (2001).

\bibitem{Les Houches}
Derrida, B.,
\textit{Lecture notes on large deviations in non-equilibrium diffusive systems},
SciPost Physics Lecture Notes, 106 (2025).

\bibitem{Espigares_Hurtado-Garrido-PRL-(2013)}
Espigares, C.~P., Garrido, P.~L., Hurtado, P.~I.,
\textit{Dynamical phase transition for current statistics in a simple driven diffusive system},
Physical Review E \textbf{87}, 032115 (2013).

\bibitem{Hurtado-(2025)}
Hurtado, P.~I.,
\textit{Optimal paths and dynamical symmetry breaking in the current fluctuations of driven diffusive media},
SciPost Physics Lecture Notes, 121 (2026).

\bibitem{BDGJL-2005-PRL}
Bertini, L., De Sole, A., Gabrielli, D., Jona-Lasinio, G., Landim, C.,
\textit{Current fluctuations in stochastic lattice gases},
Physical Review Letters \textbf{94}, 030601 (2005).

\bibitem{BDGJL-2006-JSP}
Bertini, L., De Sole, A., Gabrielli, D., Jona-Lasinio, G., Landim, C.,
\textit{Non equilibrium current fluctuations in stochastic lattice gases},
Journal of Statistical Physics \textbf{123}, 237--276 (2006).


\bibitem{Appert}
Appert-Rolland, C., Derrida, B., Lecomte, V., van Wijland, F.,
\textit{Universal cumulants of the current in diffusive systems on a ring},
Physical Review E \textbf{78}, 021122 (2008).

\bibitem{Bodineau-Derrida-2005}
Bodineau, T., Derrida, B.,
\textit{Distribution of current in nonequilibrium diffusive systems and phase transitions},
Physical Review E \textbf{72}, 066110 (2005).

\bibitem{Zarfaty}
Zarfaty, L., Meerson, B.,
\textit{Statistics of large currents in the Kipnis--Marchioro--Presutti model in a ring geometry},
Journal of Statistical Mechanics: Theory and Experiment \textbf{2016}, 033304 (2016).




\bibitem{Antoine}
Derrida, B., Gerschenfeld, A.,
\textit{Current fluctuations in one dimensional diffusive systems with a step initial density profile},
Journal of Statistical Physics \textbf{137}, 978--1000 (2009).

\bibitem{Berlioz}
Berlioz, T., B\'enichou, O., Grabsch, A.,
\textit{Tracer and current fluctuations in driven diffusive systems},
Physical Review Letters \textbf{134}, 247101 (2025).

\bibitem{Bet1}
Bettelheim, E., Smith, N.~R., Meerson, B.,
\textit{Full statistics of nonstationary heat transfer in the Kipnis--Marchioro--Presutti model},
Journal of Statistical Mechanics: Theory and Experiment \textbf{2022}, 093103 (2022).


\bibitem{Krapivsky}
Krapivsky, P.~L., Mallick, K., Sadhu, T.,
\textit{Tagged particle in single-file diffusion},
Journal of Statistical Physics \textbf{160}, 885--925 (2015).

\bibitem{Mallick}
Mallick, K., Moriya, H., Sasamoto, T.,
\textit{Exact solution of the macroscopic fluctuation theory for the symmetric exclusion process},
Physical Review Letters \textbf{129}, 040601 (2022).

\bibitem{Gra1}
Grabsch, A., Venturelli, D., B\'enichou, O.,
\textit{Macroscopic fluctuation theory of interacting Brownian particles},
Physical Review E \textbf{113}, 054128 (2026).

\bibitem{Akkermans-Bodineau-Derrida-Shpielberg}
Akkermans, E., Bodineau, T., Derrida, B., Shpielberg, O.,
\textit{Universal current fluctuations in the symmetric exclusion process and other diffusive systems},
EPL (Europhysics Letters) \textbf{103}, 20001 (2013).

\bibitem{Bodineau-2026}
Bodineau, T., Derrida, B.,
\textit{System driven out-of equilibrium by weak contacts with reservoirs},
arXiv:2605.01900 (2026).

\bibitem{DL}
Derrida, B., Lebowitz, J.~L.,
\textit{Exact large deviation function in the asymmetric exclusion process},
Physical Review Letters \textbf{80}, 209--213 (1998).

\bibitem{Touchette}
Touchette, H.,
\textit{The large deviation approach to statistical mechanics},
Physics Reports \textbf{478}, 1--69 (2009).

\bibitem{Touchette-bis}
Burenev, I.~N., Cloete, D.~W.~H., Kharbanda, V., Touchette, H.,
\textit{An introduction to large deviations with applications in physics},
SciPost Physics Lecture Notes, 104 (2025).

\bibitem{Touchette-ter}
Chetrite, R., Touchette, H.,
\textit{Nonequilibrium Markov processes conditioned on large deviations},
Annales Henri Poincar\'e \textbf{16}, 2005--2057 (2015).

\bibitem{DS}
Derrida, B., Sadhu, T.,
\textit{Large deviations conditioned on large deviations I: Markov chain and Langevin equation},
Journal of Statistical Physics \textbf{176}, 773--805 (2019).

\bibitem{Derrida-doucot-Roche}
Derrida, B., Dou\c{c}ot, B., Roche, P.-E.,
\textit{Current fluctuations in the one-dimensional symmetric exclusion process with open boundaries},
Journal of Statistical Physics \textbf{115}, 717--748 (2004).

\bibitem{derrida1983}
Derrida, B.,
\textit{Velocity and diffusion constant of a periodic one-dimensional hopping model},
Journal of Statistical Physics \textbf{31}, 433--450 (1983).

\bibitem{Bodineau-Derrida-2007}
Bodineau, T., Derrida, B.,
\textit{Cumulants and large deviations of the current through non-equilibrium steady states},
Comptes Rendus Physique \textbf{8}, 540--555 (2007).

\bibitem{KOV}
Kipnis, C., Olla, S., Varadhan, S.~R.~S.,
\textit{Hydrodynamics and large deviation for simple exclusion processes},
Communications on Pure and Applied Mathematics \textbf{42}, 115--137 (1989).

\bibitem{Goncalves}
Gon\c{c}alves, P.,
\textit{Hydrodynamics for symmetric exclusion in contact with reservoirs},
dans Stochastic Dynamics Out of Equilibrium,
Springer Proceedings in Mathematics \& Statistics \textbf{282},
137--205 (2019).

\bibitem{De Masi-Presutti-Tsagkarogiannis-Vares}
De Masi, A., Presutti, E., Tsagkarogiannis, D., Vares, M.~E.,
\textit{Current reservoirs in the simple exclusion process},
Journal of Statistical Physics \textbf{144}, 1151--1170 (2011).

\bibitem{Miron}
 A. Miron, D. Mukamel, H. A. Posch, 
\textit{Attraction and condensation of driven tracers in a narrow channel},
Physical Review E \textbf{104}, 024123 (2021)

\bibitem{Miron2}
A. Miron, D. Mukamel, H.A. Posch,
\textit{Phase transition in a 1D driven tracer model},
Journal of Statistical Mechanics: Theory and Experiment \textbf{6}, 063216 (2020)


\bibitem{Hurtado-Garrido-PRL-(2009)}
Hurtado, P.~I., Garrido, P.~L.,
\textit{Test of the additivity principle for current fluctuations in a model of heat conduction},
Physical Review Letters \textbf{102}, 250601 (2009).

\bibitem{Hurtado-Garrido-PRE-(2010)}
Hurtado, P.~I., Garrido, P.~L.,
\textit{Large fluctuations of the macroscopic current in diffusive systems: A numerical test of the additivity principle},
Physical Review E \textbf{81}, 041102 (2010).

\bibitem{giardina}
Carinci, G., Franceschini, C., Frassek, R., Giardin\`a, C., Redig, F.,
\textit{Large deviations and additivity principle for the open harmonic process},
Communications in Mathematical Physics \textbf{406}, 103 (2025).

\bibitem{Saito-Dhar}
Saito, K., Dhar, A.,
\textit{Additivity principle in high-dimensional deterministic systems},
Physical Review Letters \textbf{107}, 250601 (2011).

\bibitem{Baek1}
Baek, Y., Kafri, Y., Lecomte, V.,
\textit{Dynamical symmetry breaking and phase transitions in driven diffusive systems},
Physical Review Letters \textbf{118}, 030604 (2017).

\bibitem{Baek2}
Baek, Y., Kafri, Y., Lecomte, V.,
\textit{Dynamical phase transitions in the current distribution of driven diffusive channels},
Journal of Physics A: Mathematical and Theoretical \textbf{51}, 105001 (2018).

\bibitem{Hurtado-Garrido-PRL-(2011)}
Hurtado, P.~I., Garrido, P.~L.,
\textit{Spontaneous symmetry breaking at the fluctuating level},
Physical Review Letters \textbf{107}, 180601 (2011).

\bibitem{Shpielberg-Yaroslav-Akkermans}
Shpielberg, O., Don, Y., Akkermans, E.,
\textit{Numerical study of continuous and discontinuous dynamical phase transitions for boundary-driven systems},
Physical Review E \textbf{95}, 032137 (2017).

\bibitem{Gillespie1977}
Gillespie, D.~T.,
\textit{Exact stochastic simulation of coupled chemical reactions},
The Journal of Physical Chemistry \textbf{81}, 2340--2361 (1977).

\bibitem{Groth-Tworzydlo-Beenakker-2008}
Groth, C.~W., Tworzyd{\l}o, J., Beenakker, C.~W.~J.,
\textit{Electronic shot noise in fractal conductors},
Physical Review Letters \textbf{100}, 176804 (2008).

\bibitem{Lecomte-Imparato-vanWijland}
Lecomte, V., Imparato, A., van Wijland, F.,
\textit{Current fluctuations in systems with diffusive dynamics, in and out of equilibrium},
Progress of Theoretical Physics Supplement \textbf{184}, 276--289 (2010).

\bibitem{Spohn book}
Spohn, H.,
\textit{Large scale dynamics of interacting particles},
Springer-Verlag (1991).

\end{thebibliography}
\end{document}